\documentstyle[aps,prl]{revtex}
\newcommand{\beq}{\begin{equation}}
\newcommand{\eeq}{\end{equation}}

\newcommand{\etal}{{\it et al.}\ }

\newcommand{\tc}{$T_c$ }
\newcommand{\Mfl}{$M_{\rm fl}$ }
\newcommand{\Hup}{$H_{\rm up}$ }

\begin{document}
\def\gsim{\huge ^{_>}_{^\sim}}


\title{Superconducting Diamagnetic Fluctuations in MgB$_2$}


\author{
A.~Lascialfari$^1$,
T.~Mishonov$^{2,3}$,
A.~Rigamonti$^1$,
P.~Tedesco$^1$, and
A.~Varlamov$^4$
}


\address{%
 $^1$Department of Physics ``A.~Volta'' and Unita' INFM, University of Pavia,\\
     Via Bassi 6, I-27100 Pavia, Italy\\
 $^{2}$Laboratorium voor Vaste-Stoffysica en Magnetisme,
     Katholieke Universiteit Leuven,\\
 Celestijnenlaan 200 D, B-3001 Leuven, Belgium\\
 $^3$Department of Theoretical Physics, Faculty of Physics, Sofia University St. Kliment Ohridski,\\
     5 J. Bourchier Blvd., Bg-1164 Sofia, Bulgaria\\
 $^4$Unita' INFM Tor Vergata, I-001333 Rome, Italy
         }


\maketitle


\begin{abstract}
The fluctuating diamagnetic magnetization \Mfl at constant field H as a function of temperature and the isothermal
magnetization \Mfl  vs $H$ are measured in MgB$_2$, above the superconducting transition temperature. The
expressions for \Mfl in randomly oriented powders are derived in the Gaussian approximation of local
Ginzburg-Landau theory and used for the analysis of the data. The scaled magnetization $-M_{\rm
fl} / H^{1/2}  T$   is found to be field dependent. In the limit of evanescent field the behaviour  for Gaussian
fluctuations is obeyed while for  $H\gsim  100$~Oe the field tends to suppress
the fluctuating pairs, with a
field dependence of \Mfl close to the one expected when short wavelength
fluctuations and non-local electrodynamic effects are taken into account.
Our data, besides providing the isothermal magnetization curves
for $T> T_c (0)$ in a BCS-type superconductor such as MgB$_2$, evidence
an enhancement of the fluctuating diamagnetism
which is related to the occurrence in this new superconductor of
an anisotropic spectrum of the superconducting fluctuations.
\end{abstract}
\vfill
PACS numbers:
74.40.+k,
74.30.-e,
74.70.-b,
74.20.De
\vfill





Thermodynamical fluctuations on approaching the superconducting (SC) transition temperature from above yield
the formation of evanescent SC droplets causing a bulk diamagnetic magnetization $-$\Mfl. This fluctuating
diamagnetism is strongly enhanced in high temperature superconductors with respect to conventional SC's because of
the high temperature range and of the anisotropy of the cuprates\cite{AusloosVarlamov}. In spite of the difficulty
of evidencing the  superconducting fluctuations (SF) in low-temperature conventional BCS superconductors, the
fluctuating diamagnetism (FD) can be detected also in these systems by means of SQUID magnetization measurements.
Early data for \Mfl at constant field as a function of temperature in zero dimensional limit (aluminum particles
of size less than 1000~\AA) and in metals compounds, evidenced the rounding of the transition due to SF and the
effect of the magnetic field in quenching the fluctuating Cooper pairs\cite{Tinkham}. From the measurements of
\Mfl vs $T$ of Gollub \etal\cite{Gollub} one can deduce the occurrence of an upturn in the field dependence of
$M_{\rm fl}$: for $H\ll H_{\rm up}$ the diamagnetic magnetization increases with H while for
$H \gsim H_{\rm up}$ the
field tends to suppress the fluctuating magnetization. The upturn field \Hup can be approximately related to the
Ginzburg-Landau (GL) coherence length $\xi(T)$ (see later on). Most likely in view of the small value of
\Mfl in conventional BCS superconductors, isothermal magnetization curves  \Mfl vs $H$ have not been studied in
detail, in the authors's knowledge. The relevance of the field dependence of $M_{\rm fl} (T,H)$
for the study of FD has been recently stressed in the framework of a Gaussian GL approach for non-isotropic
systems\cite{Mishonov,MishonovPenev}. The new superconductor MgB$_2$\cite{Yamashita}, although being of ``conventional'' BCS
character has two characteristics that can be expected to enhance SF and therefore the value of \Mfl at
$T=T_c(0)$: the high value of the transition temperature and, as we will deduce later on,
an anisotropic spectrum
of the fluctuations, similarly to cuprate superconductors.


This report deals with a study of FD in MgB$_2$ by
means of high resolution SQUID magnetization measurements, with a successful detection of the magnetization curves
--\Mfl vs $H$ in the temperature range  $T_c(0) \lesssim T \lesssim T_c(0) + 0.5$~K. From the experimental findings we have been
able to prove that \Mfl and its field dependence are close to the one predicted by theories based on the extension
of the GL approach to include short wavelength fluctuations of the order parameter and non local electrodynamic
effects. It is also deduced that MgB$_2$  has an anisotropic spectrum of fluctuations, with remarkable enhancement
of the FD.


The measurements have been carried out by means of the  Quantum Design MPMS-XL7  SQUID magnetometer, allowing one
to achieve  temperature resolution up to 1~mK and the measure of a lowest magnetic moment  value
around 10$^{-7}$~emu,
in practice corresponding with our samples to the detectability of \Mfl in field less than one Oersted. The
magnetic field could be increased up to 7$\times 10^4$~Oe. The sample was prepared by Palenzona \etal (University
of Genova) starting from high purity B and Mg powders, heated at 950~$^\circ$C for 24 hs. A temperature-independent
paramagnetic magnetization was detected in the range 100-40~K, yielding a volume Pauli susceptibility around
2$\times 10^{-7}$. The diamagnetic  magnetization around \tc was obtained from the raw data by
subtracting the paramagnetic magnetization, as well as the temperature-independent correction due
to the sample holder. Zero field cooled (ZFC) magnetization curves have been compared to field cooled (FC)
data, obtained by cooling to a given temperature in the presence of different fields.


In Fig.~1 the temperature dependence of the scaled volume magnetization $m=-M_{\rm fl} / H^{1/2} T_c$ is reported,
for MgB$_2$, and for comparison (see discussion later on), for optimally doped YBCO (oriented powder). One can
deduce the following. In the temperature range indicated by a), $m$ linearly decreases on increasing temperature,
corresponding to a diamagnetic susceptibility going as $(1-T/T_c(H))$ as expected\cite{Tinkham,Gollub} outside
from the critical region below $T_c$. From this behaviour one can deduce $T_{c2}(H)=T_c(H)$ at various fields (the
superconducting transition remains second order even in the presence of the field) and we derived  $(dH_{c2}/dT)$
for $H\rightarrow 0$ around 800~Oe/K. In the vicinity of $T_c(H)$ the rounding of the transition due to critical
fluctuations\cite{Gollub} originates an almost exponential temperature dependence for $m$. Because of the
broadened transition region, and possibly of the powder character of the sample, the zero field transition
temperature can be estimated only with a rather large uncertainty: $T_c=T_c (0) = 39.07\pm 0.04$~K.


Now we are going to present and analyze the isothermal magnetization curves (Fig.~2). The
departure from the Prange law\cite{Tinkham}, $M_{\rm fl}(T = T_c)\propto H^{1/2}$ and $M_{\rm fl} (T\gg T_c )
\propto H$, valid in the limit $H\rightarrow 0$, is evident. The upturn field $H_{\rm up}(T)$, as well as the
maximum negative value reached by \Mfl at $H_{\rm up}$, are functions of temperature. Let us first give a simple qualitative
justification of these experimental findings, based on the assumption that the  fluctuation-induced evanescent SC
droplets are spherical, with a diameter of the order of the coherence length $\xi(T)$. It is noted that those
droplets are the ones actually yielding the most effective diamagnetic screening\cite{Gollub}.



For the
droplets of radius $d\ll \xi(T)$ the zero dimensional approximation can be used. In this case
the order
parameter is no longer spatial dependent and for the magnetization
an exact solution in the framework of the GL functional, valid for all fields $H \ll
H_{c2}(0)$ can be found\cite{LarkinVarlamov}
\beq
M_{\rm fl}^{D=0} = -k_{\rm B} T \frac{2\pi^2 \xi(0)^2 d^2 /5 \Phi_0^2}{(\varepsilon + \pi^2 \xi(0)^2 H^2 d^2 /5 \Phi_0^2)}
\cdot H
\eeq
Now extending this result for $d\sim \xi(T)$ the upturn field turns out
$H_{\rm up} = \varepsilon\Phi_0/\xi(0)^2$, with
$\varepsilon=[(T-T_c(0))/T_c(0)]$ while the magnetization at \Hup decreases on increasing temperature
approximately as $M_{\rm fl}(H_{up})\propto\varepsilon^{-1}$. The data in Fig.~2 are  qualitatively accounted for by the above
considerations.


In Fig.~3 the value of the scaled magnetization  $m = -M_{\rm fl} (T) / H^{1/2} T$   is reported
as a function of the magnetic field. It is noted that $m_c = m (T=T_c)$  decays with the field and reaches half of
the value $m_c(H\rightarrow 0)$ at about $H_s = 100$~Oe. For temperature far from $T_c(0)$ and small field, the
magnetization tends to the linear increase with $H$ and then $m
\propto H^{1/2}$. For larger $H$ the departure  from the behaviour expected in the framework of Gaussian GL
theories in finite fields is dramatic. It can be observed that the field dependence of the FD
in MgB$_2$ is similar to the one
expected according to theories taking into account short wave length fluctuations
and non local electrodynamic effects, as detected in conventional BCS
superconductors\cite{Tinkham}.


Thus we first discuss the experimental findings on the basis of the fluctuation part of the Gibbs free
energy of an anisotropic superconductor\cite{Mishonov}
\beq
G(\varepsilon,h)=\frac{V k_BTh^{3/2}}{2^{1/2}\pi\xi^3(0)}
\zeta\left(-\frac{1}{2},\frac{1}{2}+\frac{\varepsilon}{2h}\right),\qquad
h=\frac{H}{H_{c2}(0)},\qquad
\eeq
where $\xi(0)=[\xi_{ab}^2(0)\xi_c(0)]^{1/3}$ here is the geometrical average of the three components of the coherence length
and $H_{c2}(0)$ is angular dependent.
For a randomly oriented powder
we have to average with respect to the angle $\theta$ between the magnetic field
and the $c$-axis of the microcrystals, resulting
\beq
G(\varepsilon,H)=
2\pi^{1/2}k_BTV\int_0^1\left(\frac{g(z)H}{\Phi_0}\right)^{3/2}
\zeta\left(-\frac{1}{2},\frac{1}{2}+\frac{\Phi_0\,\varepsilon}{4\pi\xi^2(0)g(z)H}\right)\,d z,
\eeq
where the anisotropy function
\beq
g(z;\gamma)\equiv\gamma^{-1/3}\sqrt{1+(\gamma^2-1)z^2},\qquad
z\equiv\cos\theta=H_z/H,
\eeq
describes the angular dependence of the upper critical field $H_{c2}\propto 1/g(\theta)$
and $\gamma=\xi_{ab}(0)/\xi_c(0)$ is the anisotropy parameter.
%
%
%
%



The averaged magnetization can be derived by differentiation of $G$ with respect to the field\cite{MishonovPenev},
yielding
\begin{eqnarray}
M_{\rm fl}(\varepsilon,h)=-\frac{1}{V}\left(\frac{\partial G}{\partial H}\right)_T
&=&\frac{3\sqrt{\pi}}{\Phi_0^{3/2}}k_BT\sqrt{H}\int_{0}^{1}g^{3/2}(z)
 \zeta\left(-\frac{1}{2},\frac{1}{2}+\frac{\Phi_0\,\varepsilon}{4\pi\xi^2(0)g(z)H}\right)\,d z\nonumber\\
&&+\frac{k_BT\varepsilon}{4\xi^2(0)\left(\pi\Phi_0H\right)^{1/2}}
   \int_{0}^{1}g^{1/2}(z)
   \zeta\left(\frac{1}{2},\frac{1}{2}+\frac{\Phi_0\,\varepsilon}{4\pi\xi^2(0)g(z)H}\right)\,d z .
\end{eqnarray}
For the magnetization at \tc one has
\begin{eqnarray}
\label{strong}
M_{\rm fl}(T_c,H)&=&C_0\frac{k_BT_c}{\Phi_0^{3/2}}J(\gamma)\sqrt{H}
\quad \rm{with}
\quad C_0=3\sqrt{\pi}\left[-1+\frac{1}{\sqrt{2}}\right]\zeta\left(-\frac{1}{2}\right)=0.32377
\\
\rm{and}
\quad
J(\gamma)&\equiv&\int_0^1g^{3/2}(z;\gamma)d z
=\gamma\int_0^1\left[\left(1-\frac{1}{\gamma^2}\right)z^2+\frac{1}{\gamma^2}\right]^{3/4}dz
\approx\frac{2}{5}\gamma,\quad\mbox{for }\gamma\gg1,
\end{eqnarray}
where $\zeta\left(-\frac{1}{2}\right)=-0.207886\dots$ is the Riemann zeta-function.
The domain of applicability of the Prange's law $M_{\rm fl}\propto \sqrt{H} T\simeq const$ can be significantly limited by
critical fluctuations, anisotropy and nonlocality effects. It can be written as
\beq
\frac{\pi\gamma^{1/3}\mu_0^2(k_BT_c)^2\lambda^4(0)}{2\Phi_0^3\xi^4(0)}
\ll H \ll
\frac{\Phi_0}{2\pi\xi^2(0)\gamma^{2/3}},\qquad \mu_0=4\pi,
\eeq
while the fluctuating  magnetization above $T_c$ is
\begin{equation}
\label{weak}
M_{\rm fl}(T>T_c,H\rightarrow 0)
 =-\frac{\pi}{6}\frac{k_BT}{\Phi_0}\frac{\xi(0)}{\sqrt{\varepsilon}}K(\gamma)H,\quad
 K(\gamma)\equiv\int_0^1g(z;\gamma)d z
 =\frac{2}{3}\gamma^{-2/3}+\frac{1}{3}\gamma^{4/3} .
\end{equation}


Let us comment on the absolute value of the fluctuating magnetization around the transition. According to
the above equations and to scaling arguments\cite{SchneiderKeller,Junod,SchneiderSinger}, one has
\beq
m_{\rm c}={{M_{\rm fl}{\rm (T_c)}}\over{\sqrt{H} \cdot {\rm T_c}}}=
{{k_{\rm B}}\over{\Phi_0^{3/2}}} \cdot m_D(\infty)={{k_{\rm B}}\over{\Phi_0^{3/2}}} \cdot(-0.324)\cdot \gamma
\eeq
%
%
%
where for anisotropic superconductors the field should be along the c-axis (cf. Ref. 4, 5).


From the data for $m$ reported in Fig.~1 one sees that the results obtained in MgB$_2$ are consistent with a
spectrum of SF with a strong enhancement factor. For a quantitative estimate, also in view of
the uncertainty in the transition temperature,  we
preferred to compare the value of $m_c$  with the one in grain-oriented YBCO placed in magnetic field along the c-axis.
It turns out that the absolute value of the scaled magnetization  is slightly field dependent also in YBCO, where
the crossing of $m (T)$ for different fields at $T_c(0)$ seems to be well verified\cite{Junod} for $H\gsim 1000$~Oe.
While our data coincide with the one in Ref.\cite{Junod} for H=1000~Oe, a significative increase of $m$  appear
when the field is reduced to 3~Oe. By using our data in small field and comparing the results for m at $T_c(0)$ in
powdered MgB$_2$ and in YBCO, where $\gamma\simeq 7$,
in the light of Eq.~(6) and (7) one deduces an anisotropic factor  of the same order in
both cases. It should be remarked that the anisotropy degree of MgB$_2$ is still uncertain,
with reports giving values unexpectedly large particularly for powders, ranging from 5 to 9
\cite{11}.



In summary, by means of SQUID magnetization measurements we have experimentally detected the fluctuating
diamagnetism and the magnetization curves above $T_c$ in MgB$_2$. The data have been first analyzed in terms of the
theory for powdered anisotropic superconductor within the Gaussian approximation of the GL scenario. It has been
found that only in the limit of evanescent field the theory is obeyed, while for
$H\gsim 100$~Oe the field tends to
suppress the fluctuating pairs and the behavior of \Mfl is similar to the one observed in conventional BCS
superconductor and attributed to short wavelengths fluctuations and non local effects. The absolute value of the
magnetization at T$_c$  indicates that the spectrum of fluctuations in MgB$_2$ is clearly anisotropic, to an extent
close to the one in optimally doped YBCO.


%
\acknowledgments
Useful discussions with R.~Gonnelli, G.~Giunchi and A.~Palenzona are gratefully
acknowledged. One of the authors (T.M.) is supported by the Belgian DWTC, IUAP, the Flemish GOA and VIS/97/01
Programmes, KUL Senior Fellow (F/00/038).



%
\begin{figure}
\label{Figure 1}
\figure{{\bf{FIG.~1}}. Temperature dependence of the scaled magnetization $m=-M_{\rm fl}/H^{1/2}T_c$ in MgB$_2$ around the
superconducting transition temperature in a field of 1~Oe (empty squares).
For comparison (see text) the data in
oriented powder of optimally doped YBCO for field along the c-axis are reported, for $H=3$~Oe (full circles) and for
$H=1000$~Oe (empty triangles), with the dotted line for the eye.
The value of the scaled magnetization expected at $T_c(0)$
for isotropic 3D systems and Gaussian fluctuations, 4.7$\times 10^{-7}$, is indicated.}
\end{figure}
%


%
\begin{figure}
\label{Figure 2}
\figure{\noindent {\bf{FIG.~2}}. ZFC isothermal magnetization curves in MgB$_2$ at typical temperatures. No difference was
observed for the magnetization data in FC condition (data not reported). While in the limit
$H\rightarrow 0$ one has -$M_{fl}(T\simeq T_c)\propto H^{1/2}$ and $-M_{fl}(T\gg T_c)\propto H$,
for field $H\gsim 100$ Oe the departure from the behavior expected in the framework of Gaussian
GL theories is dramatic. For a qualitative justification of the upturn field $H_{up}$
and of -$M_{fl}(H_{up})$ see text.}
\end{figure}


%
\begin{figure}
\label{Figure 3}
\figure{\noindent {\bf{FIG.~3}}. Field dependence of the scaled magnetizations at typical
temperatures. For $T\simeq T_c$ (squares) the Prange law $M(T_c)\propto H^{1/2}$ is verified,
for $H\lesssim 50$ Gauss. For $T\gsim T_c + 0.2K$ (triangles and circles) one has
$-M_{fl}\propto H$, again only for $H\lesssim$50 Gauss.}
\end{figure}
%


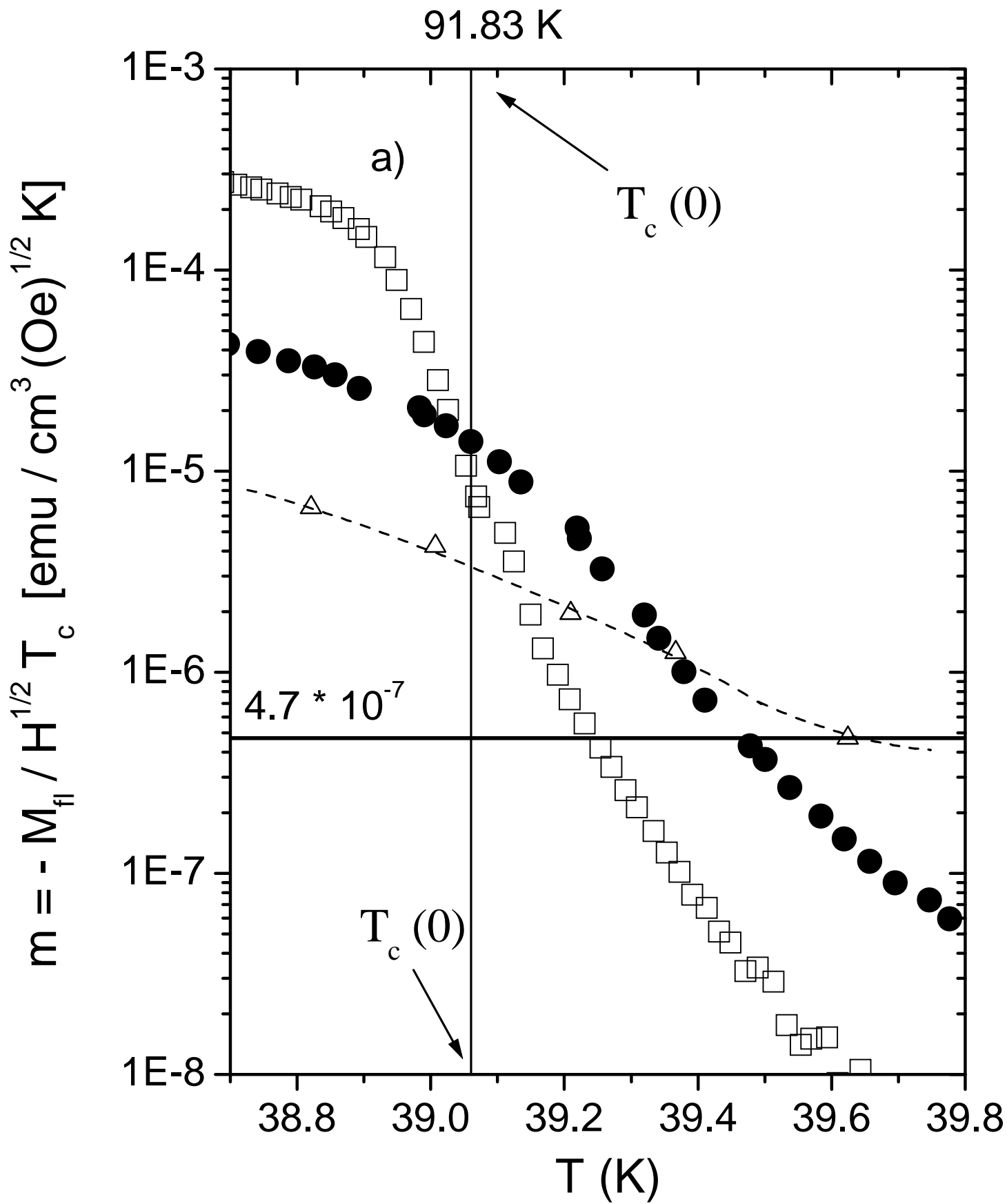
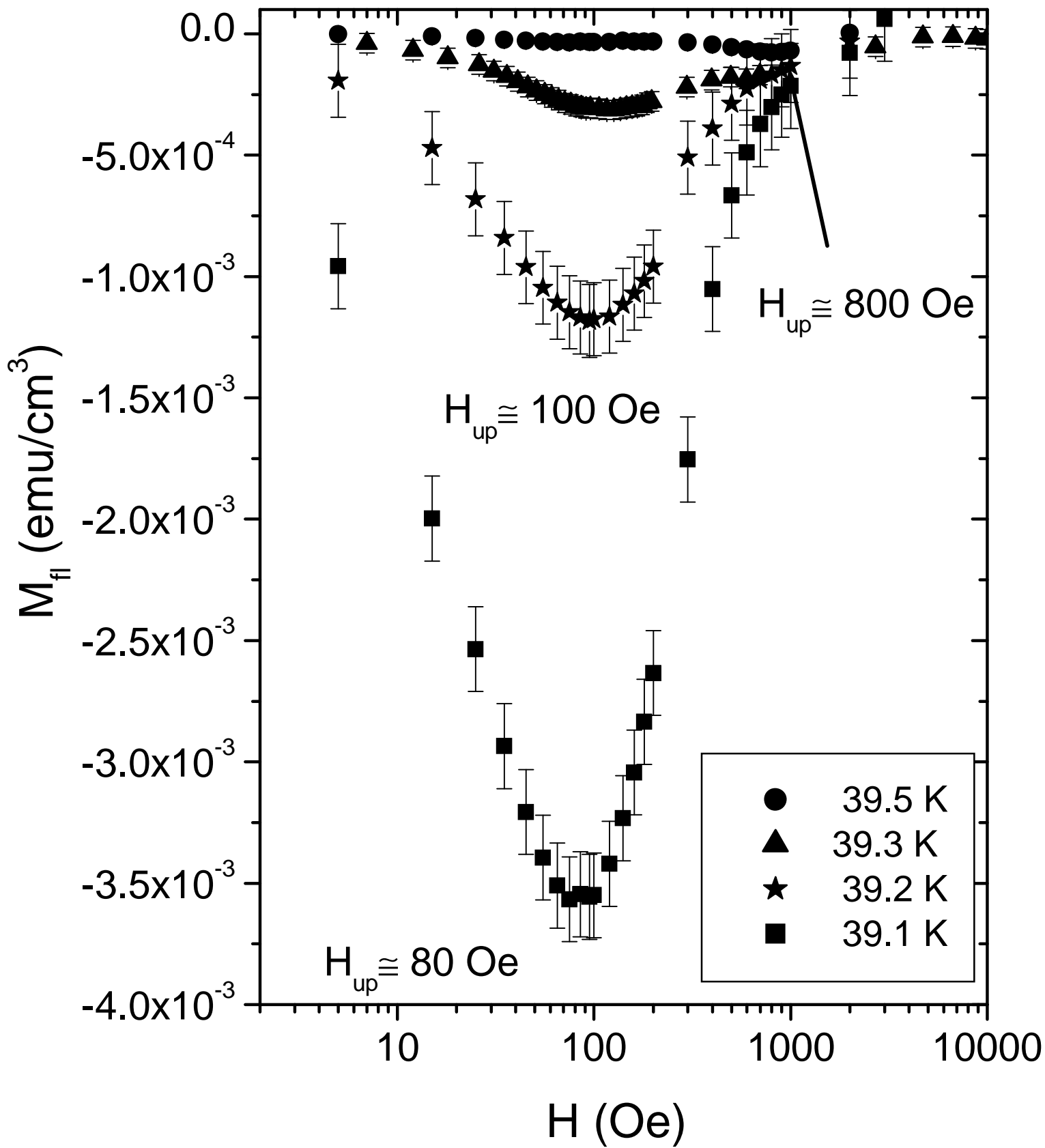
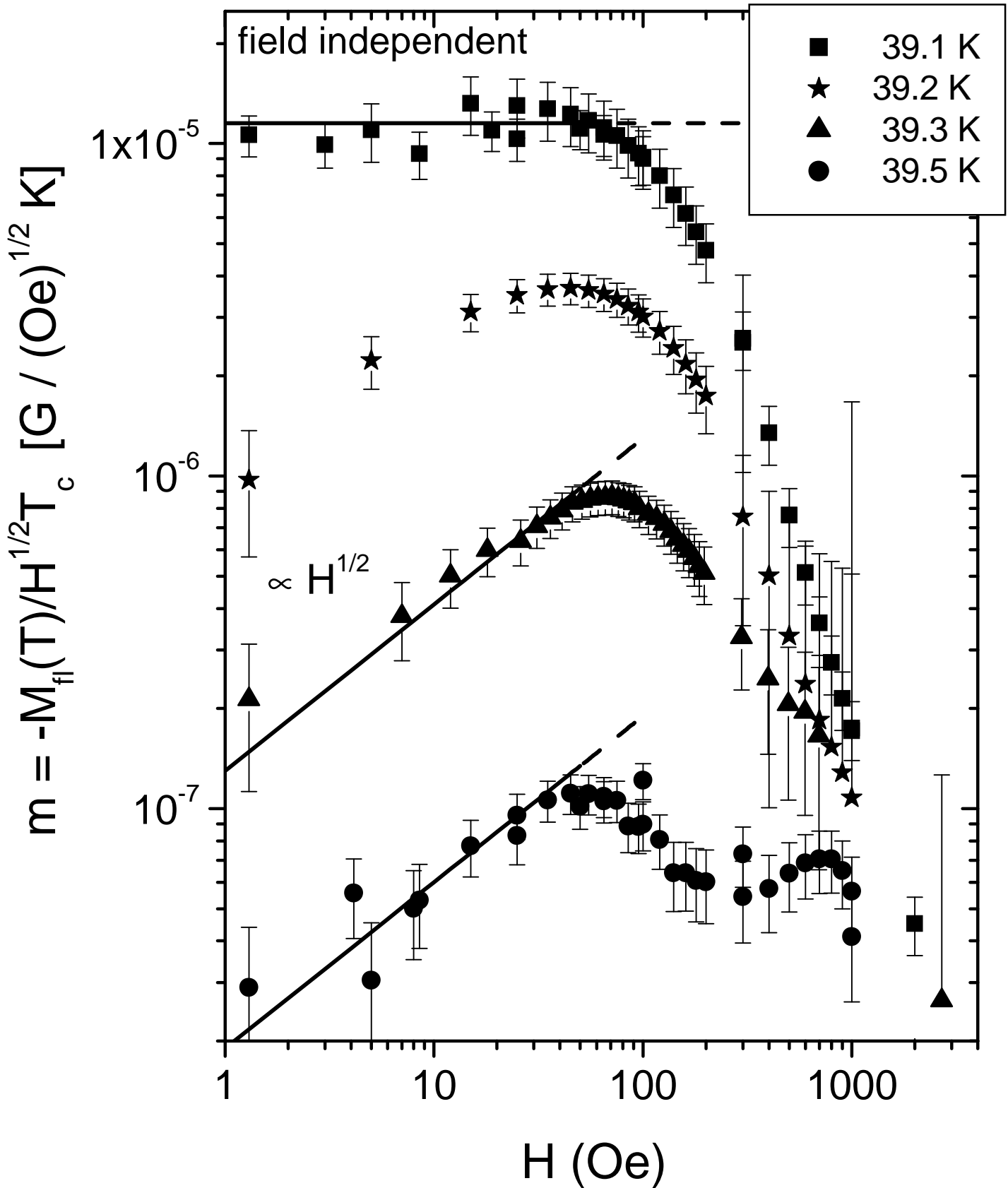
\end{document}